\documentstyle[11pt,amssymb,epsf]{article}
\makeatletter
\@addtoreset{equation}{section}
\makeatother

\def\ben{\begin{equation}}
\def\een{\end{equation}}

  \let\n=\nu  \let\p=\pi

\let\C=\Chi

\def\nn{\nonumber} \def\bd{\begin{document}} \def\ed{\end{document}}
\def\ds{\documentstyle} \let\fr=\frac \let\bl=\bigl \let\br=\bigr
\let\Br=\Bigr \let\Bl=\Bigl
\let\bm=\bibitem
\let\na=\nabla
\let\pa=\partial \let\ov=\overline
\newcommand{\be}{\begin{equation}}
\newcommand{\ee}{\end{equation}}
\def\ba{\begin{array}}
\def\ea{\end{array}}
\def\ft#1#2{{\textstyle{\frac{\scriptstyle #1}{\scriptstyle #2} } }}
\def\fft#1#2{{\frac{#1}{#2}}}
\def\del{\partial}
\def\vp{\varphi}
\def\sst#1{{\scriptscriptstyle #1}}
\def\oneone{\rlap 1\mkern4mu{\rm l}}
\def\td{\tilde}
\def\wtd{\widetilde}
\def\ie{{\it i.e.\ }}
\def\dalemb#1#2{{\vbox{\hrule height .#2pt
        \hbox{\vrule width.#2pt height#1pt \kern#1pt
                \vrule width.#2pt}
        \hrule height.#2pt}}}
\def\square{\mathord{\dalemb{6.8}{7}\hbox{\hskip1pt}}}
\newcommand{\ho}[1]{$\, ^{#1}$}
\newcommand{\hoch}[1]{$\, ^{#1}$}
\newcommand{\bea}{\begin{eqnarray}}
\newcommand{\eea}{\end{eqnarray}}
\newcommand{\ra}{\rightarrow}
\newcommand{\lra}{\longrightarrow}
\newcommand{\Lra}{\Leftrightarrow}
\newcommand{\bp}{\tilde \beta^\prime}
\newcommand{\tr}{{\rm tr} }
\newcommand{\Tr}{{\rm Tr} }
\def\0{{\sst{(0)}}}
\def\1{{\sst{(1)}}}
\def\2{{\sst{(2)}}}
\def\3{{\sst{(3)}}}
\def\4{{\sst{(4)}}}
\def\5{{\sst{(5)}}}
\def\6{{\sst{(6)}}}
\def\7{{\sst{(7)}}}
\def\8{{\sst{(8)}}}
\def\n{{\sst{(n)}}}
\def\cA{{{\cal A}}}
\def\cB{{{\cal B}}}
\def\cF{{{\cal F}}}
\def\cH{{{\cal H}}}
\def\tV{\widetilde V}
\def\tW{\widetilde W}
\def\tH{\widetilde H}
\def\tE{\widetilde E}
\def\tF{\widetilde F}
\def\tA{\widetilde A}
\def\im{{{\rm i}}}
\def\tY{{{\wtd Y}}}
\def\ep{{\epsilon}}
\def\vep{{\varepsilon}}
\def\bD{{{\bar D}}}
\def\R{{{\mathbb R}}}
\def\C{{{\mathbb C}}}
\def\H{{{\mathbb H}}}
\def\CP{{{\mathbb C}{\mathbb P}}}
\def\RP{{{\mathbb R}{\mathbb P}}}
\def\Z{{{\mathbb Z}}}
\def\bA{{{\mathbb A}}}
\def\bB{{{\mathbb B}}}
\def\bC{{{\mathbb C}}}
\def\bD{{{\mathbb D}}}
\def\bE{{{\mathbb E}}}
\def\bZ{{{\mathbb Z}}}
\def\Re{{{\frak{Re}}}}
\def\Im{{{\frak{Im}}}}
\def\cosec{{\,\hbox{cosec}\,}}
\def\Gm{{\Gamma_{\!\! -}}}
\def\Gp{{\Gamma_{\!\! +}}}
\def\stan{{standard }}
\def\nonstan{{supernumerary }}
\def\p{{\partial}}
\def\kdel#1{{\fft{\del}{\del#1}}}
\def\bG{{{\bf G}}}

\def\bog{{Bogomolny }}

\newcommand{\eq}[1]{(\ref{#1})}

\newcommand{\tamphys}{\it George and Cynthia Woods Mitchell  Institute
for Fundamental Physics and Astronomy,\\
Texas A\&M University, College Station, TX 77843, USA}

\newcommand{\auth}{
H. L\"u\hoch{\dagger\star},
M.J. Perry \hoch{\ddagger}, C.N. Pope\hoch{\dagger,\ddagger}
 and E. Sezgin\hoch{\dagger}
}

\begin{document}

\begin{flushright}
\hfill{
MIFP-08-33}\\
%\hfill{
%\bf hep-th/yymmnnn}
\end{flushright}

\vspace{25pt}

\begin{center}

{\large {\bf Kac-Moody and Virasoro Symmetries of
 Principal Chiral Sigma Models}}

\vspace{25pt}
\auth

\vspace{10pt}
\hoch{\dagger}{\tamphys}

\vspace{10pt}

\hoch{\star}{\it Division of Applied Mathematics and Theoretical Physics,\\
China Institute for Advanced Study,\\
Central University of Finance and Economics, Beijing, 100081, China
}

\vspace{10pt}

\hoch{\ddagger}{\it  DAMTP, Centre for Mathematical Sciences,
 Cambridge University,\\  Wilberforce Road, Cambridge CB3 OWA, UK}

\vspace{40pt}

\underline{ABSTRACT}
\end{center}

   It is commonly asserted that there is a $\widehat G\times G$
centreless Kac-Moody extension of the manifest $G\times G$ global
symmetry of the two-dimensional principal chiral model (PCM) for the
group manifold $G$.  Here, we show that the symmetry is in fact
larger, namely $\widehat G\times \widehat G$, the full centreless
Kac-Moody extension of the entire manifest $G\times G$ global
symmetry.  Extending previous results in the literature, we also
obtain an explicit realisation of the Virasoro-like symmetry of the
PCM, generated by $K_n=L_{n+1} - L_{n-1}$ for both positive and
negative $n$.  We show that these generators obey Sugarawara-type
commutation relations with the two commuting copies of the Kac-Moody
algebra $\widehat G$.

\vspace{15pt}

\thispagestyle{empty}

\pagebreak
%\voffset=0pt
%\setcounter{page}{1}

%
%\addtocontents{toc}{\protect\setcounter{tocdepth}{2}}

%%%%%%%%%%%%%%%%%%%%%%%%%%%%%%%%%%%%%%%%

\section{Introduction}

   String theory is probably the most successful approach to quantum
gravity now available, yet there remains much that is not
understood about it.  In perturbative string theory, one starts by
investigating how the string propagates in flat Minkowski
spacetime. This amounts to an investigation of free fields propagating
on the two-dimensional world-sheet of the physical string. One of
the organising principles of such a theory is its classical invariance
under two-dimensional Weyl transformations of the string
world-sheet. Much of what is known about string theory comes from
ensuring that this Weyl symmetry is preserved in the quantum domain.
This leads to the possibility of examining strings propagating in
backgrounds other than flat spacetime.  For example, string theory on
a product of $d$-dimensional Minkowski spacetime with a group manifold
of suitable dimension is described by an exactly-solvable principal
sigma model (PCM) with a critical Wess-Zumino term that makes the
model conformally invariant at the quantum level \cite{Gepner:1986wi}.
The world-sheet supersymmetric version of the model, namely the
spinning string on a group manifold, has also been studied
\cite{Bergshoeff:1985qr}.  Further extensions using R-R rather than
NS-NS fields to support the background have been considered in recent
years, which requires the use of the Green-Schwarz formalism.

   An example of a background of this last type that has been much
studied is $AdS_5\times S_5$ in the type IIB string.  This, however,
goes beyond the remit of the present paper since its bosonic sector is
not a group manifold.  Another example, which is a group manifold, is
$AdS_3\times S^3 \times T^4$. Although this can be considered in
either the type IIA or type IIB string theory, it is of more interest
to consider it in type IIB, since then one can rotate between the use
of RR and NS-NS flux for supporting the background.  With only NS-NS
flux non-vanishing, the superstring action contains an $SL(2,\R)\times
SU(2)$ WZW model as a subsector \cite{Pesando:1998wm}.  Turning on the
RR flux means that the coefficient of the WZ term changes, reducing to
zero if the flux is rotated into purely the RR sector.  The bosonic
sector is then precisely described by a PCM.  In order to study the
intermediate cases, it becomes of interest to consider a PCM with an
adjustable coefficient $\mu$ for the WZ term.  In fact, away from
criticality (which occurs at $\mu=1$) a straightforward redefinition
enables the model to be mapped into the pure $\mu=0$ PCM case
\cite{schwarz1}.

   There exits a vast literature on the PCM and yet, it is remarkable
that properties as basic as its symmetries have still not been fully
settled. Before stating our new results, let us briefly summarize the
known symmetries and related properties of the PCM.  Although PCMs
make an appearance in string theory, and are currently fashionable as
a consequence, they have a much longer history (see \cite{schwarz1}
for a comprehensive historical summary of the literature on hidden
symmetries in two-dimensional PCMs).

      Firstly, it is known that the PCM can be derived from the
integrability condition of first order equations
\cite{Luscher:1977rq,Zakharov:1973pp}, known as a Lax pair, that
amount to certain zero curvature conditions. This implies classical
integrability and infinitely many nonlocal conserved charges that obey
a Yangian algebra, which is essentially an enveloping algebra with a
nontrivial coproduct rule (see \cite{MacKay:2004tc} for a review).  In
addition, an infinite set of local charges with spins equal to the
exponents of the associated Lie algebra modulo the Coexeter number
exist, and they commute with each other as well as the Yangian charges
\cite{Evans:1999mj}.

        The focus of the paper will be addressing the issue of
Kac-Moody and Virasoro-like symmetries.  The PCM has a manifest global
$G_L \times G_R$ invariance in any dimensions.  Half of this
invariance is a gauge transformation which can be used to fix the
value of $g$ at some point in spacetime. The other half is a genuine
symmetry of the theory, albeit a rather trivial one. What is
remarkable however is that in two spacetime dimensions this symmetry
becomes enhanced. This enhancement was first discovered by Luscher and
Pohlmeyer \cite{Luscher:1977rq} who identified an infinite set of
conserved charges in the theory, and a collection of B\"acklund
transformations that mapped one solution of the classical equations of
motion into another. Subsequently, L. Dolan \cite{Dolan:1981fq} showed
that the modulus space of classical solutions of the PCM equations of
motion admitted an action of half of the Kac-Moody algebra $\widehat
G$.  Her work was revisited, simplified and confirmed by Devchand and
Fairlie \cite{Devchand:1981wy}.  Y.S. Wu showed that the symmetry
discovered by Dolan could be extended to a direct product of a 
full Kac-Moody algebra $\hat G$ with a Lie algebra $G$, i.e. 
$\widehat G \times G$ \cite{wu}.\footnote{Note that this Kac-Moody algebra
has no central extension, and as such it is often referred to as a loop 
algebra.  In common with much of the literature in this subject, we shall
however use the terminology of Kac-Moody algebra in this paper.}
Later, it was argued in \cite{avabab} that the symmetry is instead
described by a direct product of two commuting ``half Kac-Moody'' algebras.

   In fact, as we shall demonstrate in this paper, the Kac-Moody
symmetry of the PCM is actually larger than either the 
$\widehat G \times G$ algebra found in \cite{wu} or the product of 
two commuting ``half Kac-Moody'' algebras described in \cite{avabab}.
There is a simple argument, based on an observation by Schwarz
\cite{schwarz1}, which shows that this must be so.  It is known that a
symmetric space sigma model (SSM), in which the scalar fields of a
two-dimensional theory live on a symmetric coset space $G/H$, has a
Kac-Moody symmetry $\widehat G$ \cite{lupepo1}.  Now the PCM for a
group manifold $G$ can be equivalently viewed as an SSM for the
symmetric coset space $(G\times G)/G$, and therefore from the known
results for the SSM, it must be that the PCM has the symmetry
$\widehat G\times \widehat G$.\footnote{Schwarz actually used this
argument in reverse, observing that if one accepts the result
$\widehat G \times G$ for a PCM, then the SSM with coset $G/H$ should
have a symmetry smaller than the full $\widehat G$.  It was
subsequently shown in \cite{lupepo1} that this smaller symmetry for
the SSM, exhibited explicitly in \cite{schwarz1}, is actually
augmented by additional symmetries not found in \cite{schwarz1}, to
give the full $\widehat G$ Kac-Moody symmetry for the SSM.}  One of
the main purposes of the present paper is to give an explicit
construction of the full $\widehat G\times \widehat G$ symmetry
transformations for the PCM.  Whilst this obviously contains $\widehat
G\times G$ as a subalgebra, this is, as we shall discuss in appendix
A, different from the $\widehat G\times G$ symmetry that was claimed
in \cite{wu,schwarz1}.  We shall in addition show in Appendix A how the
product of two commuting ``half Kac-Moody'' algebras found in \cite{avabab}
is also a subalgebra of the full $\widehat G\times \widehat G$ symmetry.

   Cheng \cite{bib:cheng}, Hou and Li
\cite{bib:hou}, Li, and Hao, Hou and Li \cite{bib:li}
found evidence for the existence of some kind of Viraosoro like
symmetry acting on the classical moduli space. The subject was
re-invigorated by Schwarz in 1995 when, stimulated by the string
theoretic applications, he re-examined the whole problem, and
presented arguments in support of the $\widehat G \times G$ symmetry
proposed by Wu \cite{schwarz1}.  In addition, he found that a
particular subalgebra of the Virasoro algebra also acts on the
classical moduli space.
We shall also discuss a further Virasoro-like symmetry of the model.

\section{Kac-Moody Symmetries}

The principal chiral model starts with a field $g(x)$ which map the
spacetime with coordinates $x^\mu$ into some representation $R$ of a
Lie group $G$.  Suppose that the Lie algebra of $G$ is denoted by
${\cal G}$ and $T_i$ are the generators of $G$ is this representation,
then these generators obey the commutation relation
%%%%%
\begin{equation}
[T_i,T_j] = f_{ij}{}^k T_k
\end{equation}
%%%%%
where $f_{ij}{}^k$ are the structure constants of the group $G$.  If
it is required, any object in the Lie algebra can be decomposed into
its components by using $T_i$ are a basis, thus for example if X is
Lie algebra valued, then
%%%%%
\begin{equation}
X= X^i T_i
\end{equation}
%%%%%%
defines its components $X^i$.  From $g(x)$ we can construct a gauge
field $A$, a connection, that is a one-form that takes its values in
the Lie algebra ${\cal G}$.  Explicitly,
%%%%%
\begin{equation}
A = g^{-1}dg
\end{equation}
%%%%%
The curvature of any gauge field is given by
%%%%%%
\begin{equation}
F=dA + A\wedge A
\end{equation}
%%%%%
Thus, by the Maurer-Cartan equation, the connection is flat, $F=0$, \ie
%%%%%
\be
dA + A\wedge A=0\,.\label{maurercartan}
\ee
%%%%%

   The action for the PCM is
%%%%%
\begin{equation}
I= - \ft12 \int d^n x \, {\rm Tr}({*A}\wedge A)\label{pcmaction}
\end{equation}
%%%%%
Variation of this action with respect to $g(x)$, for which it is useful
to record the lemma
%%%%%
\be
\delta A= d\Delta g+ [A,\Delta g]\,,\qquad
   \Delta g \equiv g^{-1}\delta g\,,
\ee
%%%%%
gives the equation of motion
%%%%%
\begin{equation}
d{*A} =0\,.\label{eom}
\end{equation}
%%%%%

$A$ is a left-invariant one-form since if one considers the
transformation $g \rightarrow hg$ where $h$ is a constant element of
$G$, $A$ is invariant. The theory can be reformulated in terms of a
right-invariant one-form $\bar A = -dg g^{-1}$, so that $A$ can be
rewritten in terms of $\bar A$ by the substitution
%%%%%
\begin{equation}
A = -g^{-1}\bar Ag.
\end{equation}
%%%%%
Under the transformation
$g \rightarrow gk$ where $k$ is a constant element of $G$, $\bar A$
is invariant. The curvature $\bar F$ of $\bar A$ is again zero as can be
seen  from its definition
%%%%%
\begin{equation}
\bar F = d\bar A + \bar A\wedge\bar A
\end{equation}
%%%%%
The action for the PCM can be written in terms of $\bar A$ instead of $A$
and is
%%%%%%
\begin{equation}
I= -  \ft12 \int d^n x \, {\rm Tr}({*\bar A}\wedge \bar A)\,.
\end{equation}
%%%%%
Under variations of $g$ one finds the equation of motion
%%%%%
\begin{equation}
d{*\bar A} = 0
\end{equation}
%%%%%
which is equivalent to the original equation of motion (\ref{eom}).

   So far, this discussion of the PCM has been applicable to an
arbitrary spacetime dimension $n$.  We now specialise to the
two-dimensional case, for which the global symmetry is much larger
than the $G\times G$ of a generic dimension. In order to establish the
symmetry of the two-dimensional model, it is convenient to note that
both $F=dA + A\wedge A=0$ and the field equation $d{*A}=0$ can be
derived from the integrability condition from either of the
first-order Lax equations
%%%%
\bea
dXX^{-1} &=&\fft{t}{1-t^2}\,  {*A} + \fft{t^2}{1-t^2}\,
A\,,\label{lax1}\\
d\bar X\bar X^{-1} &=& \fft{t}{1-t^2} \, {*\bar A} +
\fft{t^2}{1-t^2}\,  \bar A\,,\label{lax2}
\eea
%%%%
where $t$ is a constant spectral parameter, and $X=X(x;t)$ and
$\bar X = \bar X(x;t)$.
Writing $d=dx^+\del_+ + dx^-\del_-$, each of the above equations
yields two differential equations known as a Lax pair.

     Taking the exterior derivative of equation (\ref{lax1}) gives
%%%%
\be
d{*A} + t (dA + A\wedge A) = 0\,.
\ee
%%%
Since this must hold for all $t$, we see that this implies
the equation of motion
$d{*A}=0$ and the flat curvature condition $F\equiv dA+A\wedge A=0$.

       The Lax equations (\ref{lax1}) and (\ref{lax2}) can be
integrated to give solutions to $X$ and $\bar X$.  Alternatively, we
can expand the $X$ and $\bar X$ around $t=0$. Since $X$ and $\bar X$
are constants at $t=0$, they can be expanded in non-negative powers of
$t$:
%%%
\be
X(x;t) = \sum_{n\ge0} \Phi_n(x)\,  t^n\,,\qquad
\bar X (x;t) = \sum_{n\ge 0} \bar \Phi_n(x)\,  t^n\,.\label{Xexp}
\ee
%%%
 From now on we shall suppress the explicit indication of the $x$
dependence of $X$ and $\bar X$, but it will often be important to
indicate their $t$ dependence explicitly.  Thus we shall sometimes
write $X$ as $X(t)$.

  Substituting (\ref{Xexp}) into the Lax equation
(\ref{lax1}) and performing a
Taylor expansion in $t$, we obtain the infinite hierarchy of equations
%%%
\bea
d\Phi_0 &=&0\,,\nn\\
d\Phi_1 &=& {*A}\, \Phi_0\,,\nn\\
d\Phi_2 &=& {*A}\, \Phi_1 + A\, \Phi_0\,,\\
&\vdots& \nn
\eea
%%%
Thus the functions $\Phi_0, \Phi_1, \cdots$ can be viewed as an infinite
number of auxiliary fields, satisfying
first-order coupled equations.

   The PCM action (\ref{pcmaction}) has the manifest global
$G_L\times G_R$ symmetry given by
\be
\delta g= g\,\epsilon -\bar\epsilon\,g\,,
\ee
%%%%%
where $\epsilon$ and $\bar\epsilon$ are Lie-algebra valued infinitesimal
constant
matrices.

   It has been established that in the special case of a
two-dimensional spacetime, there exists an infinite dimensional
extension of this symmetry.  Let us first consider the right action on
$g$.  The infinite-dimensional hierarchy of symmetries is obtained by
replacing the parameter $\epsilon$ by a quantity of the general form
$X \ep X^{-1}$.  To be precise, we now take $\ep$ to be $t$-dependent
(but still, of course, constant in spacetime), with an expansion in
non-positive powers of $t$:
%%%%%
\be
\ep(t) = \sum_{n\ge 0} \ep_\n\, t^{-n}\,.
\ee
%%%%%
The right-acting transformations are then given by
%%%%
\be
\delta g= g\,\oint X(t)\epsilon(t) X(t)^{-1}\, \fft{dt}{2\pi\im\, t}\,,
\label{contint}
\ee
%%%%
where the contour of integration is a small loop that encloses the
origin.  The transformation parameter $\ep_\0$ describes the original
Lie-algebra symmetry, while the higher parameters $\ep_\n$ with $n\ge
1$ describe the hierarchy of additional symmetries.

   The contour integration in (\ref{contint}) is simply serving the
purpose of extracting the $t^0$ terms in the Laurent expansion of
$X(t)\epsilon(t) X(t)^{-1}$.  This gives us the infinite hierarchy of
symmetry transformations, with an independent Lie-algebra valued
parameter $\ep_\n$ at each level $n$.  In practice, it is generally
more convenient to re-express the transformations, as was done in
\cite{schwarz1}, in the form
%%%%%
\be
\delta(\ep,t) g = g\eta\,,\qquad \eta(t)= X(t)\ep X(t)^{-1}\,,
\label{deltag}
\ee
%%%%%
where $\ep$ is taken to be independent of $t$, and
%%%%%
\be
\delta(\ep,t) = \sum_{n\ge 0} t^n\, \delta_\n(\ep)\,.
\ee
%%%%
By equating the coefficients of a given power of $t$ on the two sides
of (\ref{deltag}) (bearing in mind that $g$ itself is independent of
$t$), we read off the $n$'th level symmetry transformation of $g$,
with Lie-algebra valued parameter $\ep$.\footnote{Although this way of
writing the $n$'th level transformation is quite convenient, one
should be careful not to be misled by the slightly ``informal''
notation.  In particular, it should be emphasised that the Lie-algebra
valued parameter for the transformation at level $n$ can be chosen
independently of the parameter at level $m$, for all $n\ne m$.}

   It is straightforward to verify that
%%%
\be
\delta A = \fft{1}{t} \, {*d}\eta\,,
\ee
%%%
it follows that the equation of motion $d{*A}=0$ is invariant under
the above transformation.

   To show that the Lax equation
(\ref{lax1}) is also invariant, and to obtain the transformation
for $X$, let us consider the following general transformation rule
%%%%
\be
\delta_1 X_2  = U X_2\,.\label{Utrans}
\ee
%%%
Here, we need to distinguish between the spectral parameter used in
the expansion of the transformations $\delta(\ep,t)$ and the spectral
parameter in $X(t)$.  We call these $t_1$ and $t_2$ respectively, and
use the notation
%%%%%
\be
\delta_1 \equiv \delta(\ep_1,t_1)\,,\qquad \hbox{and}\qquad
  X_2\equiv X(t_2)\,.
\ee
%%%%%
Under the transformation (\ref{Utrans}),
the Lax equation (\ref{lax1}) becomes
%%%%
\be
dU + [U, dX_2 X_2^{-1}] = \fft{t_2}{1-t_2^2}\,  {*\delta A} +
                          \fft{t_2^2}{1-t_2^2}\,  \delta A\,.
\ee
%%%%
This is a first-order differential equation for $U$, which has
a one-parameter family of solutions given by
%%%%
\be
U=U_0 + \fft{t_2}{t_1-t_2} \eta_1\,,
\ee
%%%%
where $\eta_1= X_1 \ep_1 X_1^{-1}$ and
$U_0$ satisfies the homogeneous equation $dU_0 +
[U_0, dX_2 X_2^{-1}]=0$, which gives
%%%
\be
U_0= X_2\epsilon_0 X_2^{-1}\,,
\ee
%%%%
where $\ep_0$ is a Lie-algebra valued parameter, arising as a constant
of integration.  Thus the transformation associated with $U_0$ acts
only on the (auxiliary) field $X$, but not the original PCM field $g$.
By requiring that the Lax equation (\ref{lax2}) be invariant under the
transformation, we can also obtain the transformation rule for $\bar
X$.  To summarise, the complete transformation of the right action is
given by
%%%:wq
\bea
\delta g &=& g\eta\,,\qquad \eta = X\epsilon X^{-1}\,,\qquad
\tilde\delta g=0\,,\nn\\
\delta_1 X_2 &=& \fft{t_2}{t_1-t_2} (\eta_1 X_2 - X_2\epsilon_1)
\,,\qquad
\tilde\delta_1 X_2 = \fft{t_1t_2}{1-t_1t_2} X_2 \, \epsilon_1
\,,\nn\\
\delta_1 \bar X_2 &=& \fft{t_1t_2}{t_1t_2-1}
g\eta_1 g^{-1} \bar X_2\,,\qquad \td \delta \bar X_2=0\,,\label{ltrans}
\eea
%%%%%
Here the $\td \delta$ transformation is the homogeneous one associated
with $U_0$, and it acts only on $X$.  We also have used $U_0$ in
$\delta_1 X_2$ transformation so as to remove the pole at $t_1=t_2$,
by choosing $\ep_0 = -t_2/(t_1-t_2)\, \ep_1$.  (Note that since all
expansions in the spectral parameters are performed around the origin,
the only pole of concern to us is that associated with the $t_1-t_2$
denominator.)

        Analogously, we can obtain the transformation rules for the
left action, given by
%%%
\bea
\bar \delta g &=& -\bar \eta g\,,\qquad \bar \eta = \bar X\bar\epsilon \bar
X^{-1}\,,\qquad
\tilde{\bar \delta} g=0\,,\nn\\
\bar \delta_1 \bar X_2 &=&
\fft{t_2}{t_1-t_2} (\bar\eta_1 \bar X_2 -
\bar X_2\bar\epsilon_1) \,,\qquad
\tilde {\bar \delta}_1 \bar X_2 =
\fft{t_1t_2}{1-t_1t_2} \bar X_2 \bar\epsilon_1
\,,\nn\\
\bar \delta_1 X_2 &=& \fft{t_1 t_2}{t_1t_2-1}
g^{-1}\bar \eta_1 g X_2\,,\qquad \tilde {\bar \delta}_1 X_2=0\,.
\eea
%%%%%

      Having obtained the complete set of transformation rules, it
is straightforward to calculate their commutators.
The commutators for $\delta$ and $\td \delta$ are given by
%%%%
\bea
[\delta_1, \delta_2] &=& \fft{t_1}{t_1-t_2} \, \delta(\ep_{12},t_1) -
\fft{t_2}{t_1-t_2} \, \delta(\ep_{12}, t_2)\,, \nn\\
{[}\delta_1, \td\delta_2{]} &=&\fft{t_1t_2}{1-t_1t_2}
\, \delta(\ep_{12},t_1) + \fft{1}{1-t_1t_2} \,
\td\delta(\ep_{12},t_2)\,,\nn\\
{[}\td\delta_1,\td\delta_2{]} &=&
\fft{t_2}{t_1-t_2}\, \td\delta(\ep_{12},t_1) - \fft{t_1}{t_1-t_2}
\, \td \delta(\ep_{12},t_2)\,,\label{rhcom}
\eea
%%%%
where $\ep_{12}\equiv [\ep_1,\ep_2]$.  The commutators for the barred
transformations $\bar\delta$ and $\td {\bar \delta}$ take the
identical form. Furthermore, the barred and unbarred transformations
commute.

      We may now consider the mode expansions of these
transformations.  Writing the Lie-algebra valued parameter $\ep$ as
$\ep=\ep^i\, T_i$, where $T^i$ are the generators of the Lie algebra,
satisfying $[T_i,T_j] = f_{ij}{}^k\, T_k$, we write\footnote{We remind
the reader at this point about the cautionary remark about the
labeling of parameters in the previous footnote.}
%%%%
\bea
\delta(\epsilon,t) &=& \sum_{n=0}^\infty \ep_i\, J^i_n\, t^n\,,\qquad
\td\delta(\epsilon,t)=\sum_{n=1}^\infty \ep_i\, J^i_{-n}\, t^n\,,\nn\\
%%%
\bar \delta(\epsilon,t)&=&\sum_{n=0}^\infty \bar\ep_i\,
  \bar J^i_{-n}\, t^n\,,\qquad
\td{\bar \delta}(\epsilon,t)=
\sum_{n=1}^\infty \bar\ep_i\, \bar J^i_{n} \, t^n\,.\label{KMmodes}
\eea
%%%%
Substituting these expansions into the commutators of the transformations,
we obtain the algebra
%%%%
\be
[J^i_m,J^i_n] = f^{ij}{}_{k}\, J^{k}_{m+n}\,,\qquad
[\bar J^i_m,\bar J^i_n]=f^{ij}{}_{k}\, \bar J^{k}_{m+n}\,,\qquad
[J^i_m, \bar J^j_n]=0\,.\label{KKcom}
\ee
%%%
This is the full Kac-Moody algebra $\widehat G_L \times \widehat G_R$
(\ie two commuting copies of the Kac-Moody extension $\widehat G$ of
the Lie algebra $G$).

   The Kac-Moody extension of the original $G_L\times G_R$ Lie algebra
symmetries that we have obtained here is larger than the commonly
claimed result $\widehat G\times G$.\footnote{It is not clear
precisely what is meant in the literature by $\widehat G\times G$; in
particular, whether the Kac-Moody extended factor $\widehat G$ is
being associated specifically with left-acting transformations or with
right-acting transformations, or neither.  Clearly the correct answer
must be completely symmetric between left and right.  Since in any
case the $\widehat G\times G$ claim is incorrect (as demonstrated by
the fact that a PCM is a special case of an SSM), establishing a
precise interpretation of the wrong claims is inessential.}  It is also
larger than the two commuting ``half Kac-Moody'' algebras found 
in \cite{avabab}.  It is
useful, therefore, to make a comparison between our result and the
earlier results in the literature.  We shall do this in appendix
\ref{compare}.

\section{Virasoro Symmetry}\label{virasec}

\underline{Virasoro transformation}

   The Kac-Moody symmetries that we obtained in the previous section were an
infinite-dimensional extension of the manifest $G_L\times G_R$ Lie
algebra symmetries of the PCM.  There are also additional
infinite-dimensional symmetries with parameters that are singlets
under the original $G_L\times G_R$, and it is to these that we now
turn.  These algebra of these symmetries is closely related to the
centreless Virasoro algebra.

   We begin by discussing the realisation of the symmetries on $g$.
First, consider the transformation
%%%%%
\be
\delta'\, g= g\, \xi'\,,\qquad \hbox{where}\qquad \xi'(t) \equiv
  -(1-t^2) \dot X(t)\, X(t)^{-1}\,.
\ee
%%%%%
A straightforward calculation shows that this implies
%%%%%
\be
\delta'\, (d{*A}) = -A\wedge A\,.
\ee
%%%%%
Although the non-zero right-hand side means that $\delta'$ is not a
symmetry transformation, we see that the right-hand side is
independent of $t$.  We can therefore obtain a symmetry transformation
by subtracting out a $\delta'(t)$ transformation at any fixed value of
$t$.  We shall take this subtraction to be at $t=0$, and define
$\delta^V(t) \equiv\delta'(t) -\delta'(0)$.  Thus we define
%%%%%
\be
\xi(t) \equiv
  -(1-t^2) \dot X(t)\, X(t)^{-1} + I\,,\qquad
\hbox{where}\qquad I\equiv \dot X(0) X(0)^{-1}\,,
\ee
%%%%%
and the symmetry transformation is
%%%%%
\be
\delta^V\, g = g\, \xi\,.
\ee
%%%%%
Note that the subtraction using $I$ removes the $t^0$ term in the
Taylor expansion of $\delta'(t)$.

   Further, independent, symmetry transformations can also be found,
given by
%%%%%
\be
\bar\delta^V\, g = - \bar\xi\, g\,,\qquad
  \hbox{where}\qquad \bar\xi \equiv
   = -(1-t^2) \dot {\bar X}(t) \bar X(t)^{-1} +
\bar I\,,
\ee
%%%%%
with
%%%%%
\be
  \bar I \equiv \dot {\bar X}(0) \bar X(0)^{-1}\,.
\ee
%%%%%
(We have already performed the analogous zero-mode subtraction.)  It
is straightforward to see that $\bar\delta^V\, (d{*\bar A})=0$, and
that therefore the $\bar\delta^V$ transformations are also symmetries
of the equations of motion.

     There is also a zero-mode symmetry transformation, given by
%%%%%
\be
\delta_\0^V \, g = -(g\, I -\bar I\, g) \equiv g\zeta \equiv
-\bar\zeta g \,.
\ee
%%%%%

   Proceeding in a similar fashion to the Kac-Moody calculations,
seeking transformations that leave the Lax equations invariant, we may
now determine the transformation rules for the fields $X$ and $\bar
X$, finding
%%%%
\bea
\delta_1^V X_2 &=& \fft{t_2}{t_1-t_2} \xi_1 -
\fft{t_1}{t_1-t_2} \xi_2\,, \qquad
\delta_1^V \bar X_2 = \fft{t_1t_2}{t_1t_2-1}
(g\xi_1 g^{-1} - \bar \xi_2 - \bar \zeta)\bar X_2\,,\nn\\
\bar \delta_1^V \bar X_2 &=& \fft{t_2}{t_1-t_2}
\bar\xi_1 -
\fft{t_1}{t_1-t_2} \bar\xi_2\,,\qquad
\bar \delta_1^V  X_2 = \fft{t_1 t_2}{
t_1t_2-1}
(g^{-1}\bar \xi_1 g - \xi_2 - \zeta) X_2\,,\nn\\
\delta^V_\0 X&=& \xi X\,,\qquad \qquad\qquad \qquad\qquad
\delta^V_\0 \bar X=\bar \xi \bar X\,.
\eea
%%%%%%

   After some tedious algebra, we find that the commutators of the
Virasoro-type transformations are given by
%%%
\bea
[\delta^V_1, \delta^V_2] &=&
\fft{t_1t_2}{t_1-t_2}\Big( \del_1(\fft{1-t_1^2}{t_1}
\delta_1^V) + \del_2(\fft{1-t_2^2}{t_2} \delta_2^V)\Big) -
\fft{2t_1t_2}{(t_1-t_2)^2} \Big(
\fft{1-t_1^2}{t_1} \delta_1^V - \fft{1-t_2^2}{t_2} \delta_2^V
\Big)\,,\nn\\
{[}\delta^V_1, \bar\delta_2^V{]}&=&
\fft{t_1t_2}{t_1t_2 -1}
\Big(\del_{t_2} (1-t_2^2) \bar\delta_2^V) -
     \del_{t_1} ((1-t_1^2) \delta_1^V)\Big)\nn\\
&&+\fft{t_1t_2 + 1}{(t_1t_2 -1)^2}
\Big(t_2 (1-t_1^2)\delta^V_1 -t_1 (1-t_2^2)
\bar\delta^V_2\Big)\nn\\
&&+ \fft{t_1t_2}{(t_1t_2 -1)^2}
\Big((t_1-t_2)(t_1t_2 - 3)
\Big)\delta_0^V\nn\\
{[}\bar \delta^V_1, \bar\delta_2^V{]} &=&
\fft{t_1t_2}{t_1-t_2}
\Big( \del_1(\fft{1-t_1^2}{t_1}
\bar\delta_1^V) + \del_2(\fft{1-t_2^2}{t_2}
\bar\delta_2^V)\Big) -
\fft{2t_1t_2}{(t_1-t_2)^2} \Big(
\fft{1-t_1^2}{t_1} \bar\delta_1^V -
\fft{1-t_2^2}{t_2} \bar\delta_2^V
\Big)\,,\nn\\
{[}\delta_0^V, \delta^V] &=& -(1-t^2) \dot \delta^V +
(t+\fft{1}{t}) \delta^V + t\, \delta_0^V\nn\\
{[}\delta_0^V, \bar\delta^V] &=&
-(1-t^2) \dot{\bar\delta}^V + (t + \fft{1}{t})
\bar\delta^V + t \delta_0^V
\eea
%%%%%

    We can also calculate the commutators of the Virasoro-type
transformations with the Kac-Moody transformations.  For these, we
find
%%%%
\bea
{[}\delta^V_0, \delta{]} &=& - (1-t^2) \dot \delta + \fft{1}{t}
(\delta - \delta(0)) + t\delta\,,\nn\\
{[}\delta^V_0, \td\delta{]} &=&  t \delta(0) +
\del_{t} \Big( (t^2 - 1)\td\delta\Big) -
\fft1{t} (t^2 - 1)\td\delta\,,\nn\\
{[}\delta_1^V, \delta_2{]} &=& \fft{t_1(1-t_2^2)}{t_1-t_2}
\dot\delta_2 + \fft{t_2(1-t_1^2)}{(t_1-t_2)^2}(\delta_2-\delta_1)
-\fft{1}{t_2} (\delta_2-\delta(0))\,,   \\
{[}\delta^V_1, \td \delta_2{]} &=& \fft{t_2}{(1-t_1t_2)^2}
(1-t_1^2) \delta_1 -  t_2 \delta(0)\nn\\
&&+\fft{t_1t_2}{1-t_1t_2} \del_2 \Big(
(t_2^2 - 1)\td\delta_2\Big) +
\fft{t_1^2t_2}{(1-t_1t_2)^2}
(t_2^2 - 1)\td\delta_2\,,\nn\\
{[}\delta_1^V, \bar \delta_2{]} &=&
\fft{t_1t_2}{t_1t_2-1}\del_2 \Big((1-t_2^2) \bar\delta_2\Big) -
\fft{t_1^2t_2}{(t_1t_2-1)^2} (1-t_2^2) \bar\delta_2 +
\fft{(1-t_1^2)t_2}{(1-t_1t_2)^2} \tilde {\bar \delta_1}\,,\nn\\
\nn\\
{[}\delta_1^V, \td {\bar \delta}_2{]} &=&
\fft{t_1}{t_1-t_2} \del_2\Big((1-t_2^2)
\tilde {\bar \delta}_2\Big) -
\fft{t_1(t_1-2t_2)}{t_2(t_1-t_2)^2}(1-t_2^2)
\tilde {\bar \delta}_2 -
\fft{t_2}{(t_1-t_2)^2}(1-t_1^2) \tilde {\bar \delta}_2\,,\nn
\eea
%%%%%
together with the ``conjugate'' commutators where the r\^oles of
barred and unbarred transformations are exchanged:
%%%%%
\bea
{[}\delta^V_0, \bar\delta{]} &=& - (1-t^2) \dot {\bar \delta}+
\fft{1}{t} (\bar \delta - \bar \delta(0)) +
t\bar\delta\,,\nn\\
{[}\delta_0^V, \td {\bar\delta} {]} &=&  t \bar\delta(0)
+\del_{t} \Big((t^2 - 1) \tilde{\bar\delta}
\Big) - \fft{1}{t} (t^2 - 1)
\tilde{\bar\delta}\,,\nn\\
{[}\bar\delta^V_1, \bar\delta_2{]} &=&
\fft{t_1(1-t_2^2)}{t_1-t_2}
\dot{\bar \delta}_2 + \fft{t_2(1-t_1^2)}{(t_1-t_2)^2}
(\bar \delta_2-\bar \delta_1)
-\fft{1}{t_2} (\bar \delta_2-\bar \delta(0))\,,  \\
{[}\bar \delta^V_2, \td{\bar\delta}_2{]} &=&
\fft{t_2}{(1-t_1t_2)^2}
(1-t_1^2) \bar \delta_1 -  t_2 \bar \delta(0)\nn\\
&&+\fft{t_1t_2}{1-t_1t_2} \del_2 \Big(
(t_2^2 - 1)\td{\bar \delta}_2\Big) +
\fft{t_1^2t_2}{(1-t_1t_2)^2}
(t_2^2 - 1)\td{\bar \delta}_2\,,\nn\\
{[}\bar\delta^V_1, \delta_2{]} &=&
\fft{t_1 t_2}{t_1 t_2-1} \del_2 \Big((1-t_2^2)\delta_2\Big) -
\fft{t_1^2 t_2}{(1-t_1 t_2)^2} (1-t_2^2) \delta_2 +
\fft{t_2}{(1-t_1 t_2)^2} (1-t_1^2) \td \delta_1\,,\nn\\
{[}\bar \delta^V_1, \td\delta_2{]} &=&
\fft{t_1}{t_1-t_2} \del_2\Big((1-t_2^2)
\tilde {\delta}_2\Big) -
\fft{t_1(t_1-2t_2)}{t_2(t_1-t_2)^2}(1-t_2^2)
\tilde {\delta}_2 -
\fft{t_2}{(t_1-t_2)^2}(1-t_1^2) \tilde {\delta}_2\,.\nn
\eea
%%%%%

\subsection{Mode expansions}

  We now perform a mode expansion for the Virasoro-like transformations,
which is given by
%%%%
\be
\delta^V(t) = \sum_{n=1}^\infty K_n t^n\,,\qquad
\bar\delta^V(t)=\sum_{n=1}^\infty K_{-n} t^n\,,\qquad
\delta_\0^V =K_0\,.
\ee
%%%%
Substituting this, and the previous Kac-Moody mode expansion
(\ref{KMmodes}), into the various commutators of transformations, we
obtain the algebra
%%%%
\bea
[J^i_m,J^i_n] &=& f^{ij}{}_{k} J^{k}_{m+n}\,,\qquad
[\bar J^i_m,\bar J^i_n]=f^{ij}{}_{k} \bar J^{k}_{m+n}\,,\qquad
[J^i_m, \bar J^j_n]=0\,,\nn\\
{[}K_m, K_n{]}&=&(m-n)\Big(K_{m+n+1} - K_{m+n-1}\Big)\,,\nn\\
{[}K_m, J^i_n{]}&=&-n \Big( J^i_{m+n+1}-J^i_{m+n-1} \Big)\,,\nn\\
{[}K_m, \bar J^i_n{]}&=&-n \Big(\bar J^i_{m+n+1}-\bar J^i_{m+n-1}\Big)\,.
\label{algebra}
\eea
%%%%
The $K_m$ generate a subalgebra of the centreless Virasoro algebra, as
may be seen by noting that if we were to define
%%%%%
\be
K_m=L_{m+1}-L_{m-1}\,,\label{KLrel}
\ee
%%%%%
where the $L_m$ satisfy $[L_m,L_n] = (m-n) L_{m+n}$, then we would
obtain precisely the algebra given in the second line of
(\ref{algebra}).  It should be emphasised, however, that the relation
(\ref{KLrel}) cannot be inverted to express the Virasoro generators in
terms of the generators $K_m$ of the symmetry transformations we have
exhibited.  In fact, only two extra generators would be needed in
order to be able to invert (\ref{KLrel}).  If, for example, we had
symmetry transformations corresponding directly to $L_0$ and $L_1$,
then the inversion would be possible.  (Alternatively, it would
suffice to have an additional symmetry associated with an $L_n$ for
any even $n$, and an additional symmetry associated with an $L_n$ for
any odd $n$.)

%%%%%%%%%%%%%%%%%%%%%%%%%%%%%%%

\subsection{Towards a full Virasoro symmetry}

It has been asserted by Devchand and Schiff that the Virasoro-like
symmetry studied above can actually be extended to a full centreless
Virasoro algebra \cite{devsch}, at least in the case that $G=U(N)$.
The method they employ is based on aspects of the symmetries as
transformations that act on the solution space. For the reader's
convenience, we outline very briefly the basic set up. To begin with,
one considers special solutions of the PCM equations that take the
form
\be
A^{(0)}= A(x^+) dx^+  + B(x^-) dx^-\ ,
\ee
%%%%%%
where $A(x^+)$ and $B(x^-)$ are arbitrary matrices lying in the Cartan
subalgebra of $U(N)$.  The Lax equation is then readily solved, to
yield
\be
X^{(0)} (x^+, x^-, \lambda) = e^{M(x^+, x^-, \lambda)} X_0 (\lambda)\,,
\label{sx}
\ee
where $\lambda=1/t$ and
\be
M (x^+, x^-, \lambda) = {1\over \lambda-1} \int_{x_0^+}^{x^+} A(y^+) dy^+ 
  - {1\over \lambda+1} \int_{x_0^-}^{x^-} B(y^-) dy^-\,,
\ee 
and $X_0(\lambda)$ is an unconstrained element of $G$ serving as the
initial condition. Next, it is assumed that any group element can be
factorized as $G=G_- G_+$ where (a) $G_-$ denotes the group of
analytic maps inside the contour ${\cal C}$ in the $\lambda$-plane
that is a union of two small contours centered around $\lambda=\pm 1$
such that $\lambda=0$ remains outside both of them, and approaching
the identity at $\lambda=\infty$; and (b) $G_+$ denotes the the group
of maps analytic in the region complementary to the contour ${\cal C}$
\cite{devsch}.\footnote{This rather non-trivial decomposition goes by
the name of Birkhoff factorisation.  For the case presented, a proof
that this is always in principle possible, even if not entirely
practical, is contained in some results of Pressley and Segal
\cite{preseg}.}  Using this factorization, the solution \eq{sx} is
written as
\be
X^{(0)}= S^{-1} Y\ ,
\ee
where $S^{-1}: M \rightarrow G_-$ and $Y: M \rightarrow G_+$. Then the 
element $S$ is expanded as
\be
S=\sum_{n=0}^\infty s_n (x^+, x^-) (1+\lambda)^n = 
\sum_{n=0}^\infty {\tilde s}_n (x^+,x^-) (1-\lambda)^n\ .
\ee
Finally, a general solution to the PCM is constructed as \cite{devsch}
\be
A_+ = s_0 A(x^+) s_0^{-1}\ , \qquad A_-={\tilde s}_0 B(x^-) 
{\tilde s}_0^{-1}\ .
\ee
The symmetries of the PCM are then viewed as various transformations
of the free fields $A(x^+), B(x^-), U_0(\lambda)$, i.e. as symmetries
that act on the solution space. In particular, Devchand and Schiff
argue that the Virasoro-like symmetries we exhibited explicitly in
section \ref{virasec} correspond to the reparameterisations
\be
\delta^V_m X_0(\lambda) = \epsilon_m \lambda^{m+1} X'_0(\lambda)\ ,
\ee
with the condition that the points $\lambda=\pm 1$ are held fixed
\cite{devsch}.  It is then argued that at the level of infinitesimal
symmetries the need to fix $\pm 1$ is superfluous and that the {\it
full} Virasoro algebra arises upon relaxing this condition.

  According to Devchand and Schiff, the complete Virasoro algebra acts
as a solution generating symmetry of the PCM.  They do not, however,
exhibit the action of the complete Virasoro algebra directly on the
original fields $g$ of the PCM, nor on the auxiliary fields $X$, and
as far as we are aware, there is no local way of doing so.

%%%%%%%%%%%%%%%%%%%%%%%%%%%%%%%%%%%%%%%%%%%

\section{PCM with WZ Term}

   As we mentioned in the introduction, if one considers an
AdS$_3\times S^3 \times T^4$ in the type IIB string, then by allowing
the 3-form flux to be sourced partly by the RR field and partly by the
NS-NS field, then the bosonic sector of the theory will be described
by a PCM in which the WZ term is non-vanishing, but with an adjustable
coefficient $\mu$ that can lie in the range $-1\le \mu\le 1$.  The
case $\mu=0$ corresponds to pure RR flux, while $\mu=\pm1$, the
``critical cases,'' correspond to the chiral or antichiral WZW model,
with pure NS-NS flux.

   To be more precise, the 3-form flux in the AdS$_3\times S^3\times T^4$
background can be written as
%%%%%
\be
F= \im \,(\mu+\im\, \nu)\, (\Omega_{AdS_3} + \Omega_{S^3})\,,
\ee
%%%%%
where $\Omega_{AdS_3}$ and $\Omega_{S^3}$ are the volume forms
on AdS$_3$ and $S^3$, the constants $\mu$ and $\nu$ satisfy
%%%%%
\be
\mu^2+\nu^2 =1\,,
\ee
%%%%%
and the complex 3-form $F$ is constructed from the RR and NS-NS
3-forms according to
%%%%%
\be
F= F^{RR} + \im\, F^{NS}\,.
\ee
%%%%%
The bosonic part of the superstring action contains a WZ term
constructed from the NS-NS 2-form potential $B^{NS}$.  The RR field
$F^{RR}$ couples only to fermionic terms in the action, and thus it is
absent from the bosonic sector.

   As was discussed in \cite{schwarz1}, the PCM with a WZ term where
$\mu$ takes a non-critical value can in fact be mapped by means of an
invertible field redefinition into the pure PCM case with $\mu=0$.  To
see this, we shall demonstrate the converse, namely that a pure PCM
can be mapped into a PCM with non-critical WZ term by means of an
invertible field redefinition.

   We begin with a standard PCM whose group manifold $G$ is
parameterised by $g(x)$, and so $A=g^{-1} dg$ satisfies the standard
Maurer-Cartan and field equations
%%%%%
\be
dA+A\wedge A=0\,,\qquad d{*A}=0\,.\label{cmeom}
\ee
%%%%%
We then introduce $Y(\mu)$, which is defined to obey the equation
%%%%%
\be
dY\, Y^{-1} = \fft{\mu^2}{1-\mu^2}\, {*A} + \fft{\mu}{1-\mu^2}\, A\,.
\label{Ydef}
\ee
%%%%%
 From this, we define a new group element $g'$, and gauge connection $A'$,
by writing
%%%%%
\be
g'= g Y\,,\qquad A'= {g'}^{-1} dg'\,.
\ee
%%%%%

   Clearly, from its definition, $A'$ satisfies the same Maurer-Cartan
equation as does $A$, namely
%%%%%
\be
dA' + A'\wedge A'=0\,.
\ee
%%%%%
A straightforward calculation shows that in consequence of (\ref{cmeom}),
the new connection $A'$ satisfies the equation of motion
%%%%%
\be
d{*A'} + \mu A'\wedge A'=0\,.
\ee
%%%%%
This equation can be derived by varying the action
%%%%%
\be
I_{WZ} = -\ft12 \int_{\del M} {\rm Tr}({*A'}\wedge A') -
  \fft{\mu}{3} \int_M {\rm Tr}(A'\wedge A'\wedge A')\label{pcmwzact}
\ee
%%%%%
with respect to the fundamental field $g'$.  This action is precisely
the one that describes a PCM with a WZ term.

   Note from (\ref{Ydef}) that we can generate such a model for any
value of $\mu$ except for the critical values $\mu=\pm1$ that arise in
the WZW model.

   Of course the PCM with non-critical WZ term has the same $\widehat
G \times \widehat G$ Kac-Moody symmetry as does the pure $\mu=0$ PCM.

\section{Conclusions}

   In this paper we have studied the Kac-Moody extension of the
manifest $G\times G$ global symmetry of a two-dimensional principal
chiral model based on the group manifold $G$.  We demonstrated that
the symmetry algebra is the full $\widehat G\times \widehat G$
centreless Kac-Moody extension of $G\times G$, and we obtained
explicit transformation rules for the complete algebra.  These results
go beyond those presented previously in the literature, where the smaller
symmetries $\widehat G\times G$ \cite{wu} or two commuting ``half Kac-Moody''
algebras \cite{avabab} were proposed.  
The $\widehat G\times \widehat G$ symmetry that we find is consistent with
the fact that a two-dimensional symmetric space sigma model $G/H$ has
a $\widehat G$ Kac-Moody symmetry, since a group manifold $G$ can be viewed
as the symmetric space $(G\times G)/G$, where the denominator lies in the
diagonal subgroup of $G\times G$.

\section*{Acknowledgements}

   We thank Henning Samtleben for discussions.  The research of C.N.P. 
is supported in part by DOE grant DE-FG03-95ER40917, and the research of
E.S. is supported in part by NSF grant PHY-0555575.

\appendix

\section{Comparison with Previous Literature}
\label{compare}

    Earlier literature on the infinite-dimensional symmetries of
the PCM has in common with our construction the subalgebra of
transformations $\delta g=g\eta $ given in (\ref{ltrans}).  We may
denote this right-acting ``half'' Kac-Moody algebra, which includes
the ordinary Lie algebra $G_R$ as the ``zero modes,'' by $\widehat
G_R^+$.  The earlier approaches and our approach diverge in the
attempt to find further infinite-dimensional symmetries over and above
$\widehat G_R^+$.

    In our case, we find the additional ``homogeneous''
transformations $\td\delta$, which leave $g$ inert but not $X$, and
which complete the algebra $\widehat G_R^+$ to give the full Kac-Moody
symmetry $\widehat G_R$.  Furthermore, we find that this right-acting
Kac-Moody symmetry has a completely independent left-acting
counterpart, giving in total a $\widehat G_L\times \widehat G_R$
Kac-Moody symmetry. An important feature of our results is that all
transformations, namely $\delta(t)$, $\td\delta(t)$ acting on the
right, and $\bar\delta(t)$ and $\td{\bar\delta}(t)$ acting on the
left, involve the use of spectral parameters that are expanded in
Taylor series around $t=0$.  Thus in our approach it is never necessary 
to make use of 
the Riemann-Hilbert transformation that invokes a relation
between $X(t)$ and $X(1/t)$.

   In previous works \cite{wu,schwarz1}, by contrast, it was claimed 
that the full
Kac-Moody extension of the manifest $G_L\times G_R$ global symmetry
was just $\widehat G\times G$, where the $\widehat G$ was associated
with the $\delta(t)$ transformations together with further
transformations that we shall call $\hat\delta(t)$, where
$\hat\delta(t)=\delta(1/t)$.  The expansion of
$\hat\delta(t)$ around $t=0$ is therefore equivalent to an expansion
of $\delta(t)$ around $t=\infty$.  It is not clear to us whether the
claim of a $\widehat G\times G$ symmetry carries with it any
implication that the $\widehat G$ is to be associated with the left
action, and the $G$ with the right action (or {\it vice versa}).  In
any case, clearly neither of these could be correct as the {\it full}
symmetry algebra, since there is a complete left-right symmetry in the
original PCM.

   Obviously there is, in a rather trivial sense, a genuine $\widehat
G \times G$ symmetry of the PCM, since $\widehat G \times G$ is a
subalgebra of the $\widehat G\times \widehat G$ symmetry that we have
exhibited in this paper.  This, however, is not the symmetry that is
discussed in \cite{wu,schwarz1}.  One way of seeing this is that if
one looks just at the zero-mode sector, the resulting $G\times G$
symmetry described in \cite{wu,schwarz1} is given by
%%%%%
\be 
\delta^\0 g = g\,\ep\,,\qquad \hat\delta^\0 g = -g_0\,
\hat\epsilon\, g_0^{-1}\, g\,,\label{gg0}
\ee
%%%%%
where $g_0$ is the value of $g(x)$ at some arbitrarily-selected point
$x_0$ in the two-dimensional spacetime.\footnote{Note that the
following discussion of the zero-mode sector of the algebra found in
\cite{wu,schwarz1} need not be restricted to the case of a
two-dimensional spacetime; it is equally applicable in {\it any}
spacetime dimension.}  The $\delta^\0$ transformation is a standard
right action of $G$ on the group manifold. However, the
$\hat\delta^\0$ transformation of $g(x)$ is highly non-local, since it
depends not only on the value of $g$ at the point $x$ but also on the
value of $g$ at the point $x_0$.  Although one might be tempted to
view $\hat\delta^\0$ simply as a left action of $G$ of the form
$\hat\delta^\0 =-\hat\ep'\, g$, where the constant parameter
$\hat\ep'\equiv g_0\, \hat\ep\, g_0^{-1}$ is just a conjugation of
$\hat\ep$ by the constant group element $g_0$, the fact that $g_0$
itself transforms under $\delta^\0$ and $\hat\delta^\0$ means that
$\hat\ep'$ itself transforms also.  In fact
%%%%%
\be
\delta^\0 g_0 = g_0\, \ep\,,\qquad \hat\delta^\0 g_0 = -g_0\,\hat\ep\,,
\ee
%%%%%
so at the spacetime point $x_0$ the $\hat\delta^\0$ is actually a {\it
right} translation, and not a left translation.

   The $\hat\delta^\0$ transformation defined in (\ref{gg0}) is just
one of an infinity of non-local transformation laws that one could
introduce,\footnote{For example, one could introduce transformations
involving conjugations of the parameter $\hat\ep$ by an arbitrarily
large number of group elements $g(x_i)$ at points $x_i$ in spacetime.}
but there does not appear to be a strong motivation for doing so since
they would all lie outside the class of transformations we normally
wish to consider.  Thus, the $\hat\delta^\0$ transformations in
(\ref{gg0}) amount to a somewhat artificial introduction of a
non-standard non-local $G$ symmetry that lies outside the usual
$G_L\times G_R$ locally-defined symmetry of a group-manifold sigma
model.  One striking way of seeing this is to note that if we do also
consider standard left translations, defined by
%%%%%
\be
\delta_L \, g= \ep\, g\,,
\ee
%%%%%
then a simple calculation shows that $\delta^\0$, $\hat\delta^\0$ and
$\delta_L$ all commute, and thus there is apparently a $G\times
G\times G$ symmetry in any group-manifold sigma model (in any
spacetime dimension)!

   The moral that we draw from the above discussion is that the
introduction of a preferred point $x_0$ in spacetime, with
transformation rules for fields at the point $x$ depending also on the
value of fields at the point $x_0$, leads to an unnecessary profusion
of extra non-local symmetry transformations, even at the zero-mode
level, that are not particularly germane to the original problem under
investigation.  By contrast, all the transformations of the $\widehat
G_L\times \widehat G_R$ symmetries that we have obtained in this paper
act locally on the fields $g$ and $X$.

   Finally, we consider the results presented in \cite{avabab}, in which
a symmetry algebra was obtained that is described as corresponding 
to two commuting copies
of a ``half Kac-Moody'' algebra.  Specifically, the algebra, given in equation
(21) of \cite{avabab}, has the commutation relations
%%%%%%
\bea
{[}Q_m^i, Q_n^j{]} &=& f^{ij}{}_k\, Q^k_{m+n}\,,\qquad m\,, n \ge 0\,,\nn\\
{[}Q_m^i, Q_n^j{]} &=& f^{ij}{}_k\, Q_m^k\,,\qquad m\ge0\ge n\,,\nn\\
{[}Q_m^i, Q_n^j{]} &=& f^{ij}{}_k\, (-Q_{m+n}^k + Q_m^k + Q_n^k)\,,\qquad
  m\,, n <0\,.\label{avababalg}
\eea
%%%%%
(We have adjusted the notation to match that which we are using in this paper.)
If we now define new generators $P^i_m$ by
%%%%%
\bea
P^i_m &=& Q^i_m\,,\qquad m\ge 0\,,\nn\\
P^i_m &=& Q^i_0 - Q^i_m\,,\qquad m<0\,,
\eea
%%%%%
then we see that these satisfy
%%%%%
\bea
{[} P^i_m,P^j_n{]} &=& f^{ij}{}_k\, P^k_{m+n}\,,\qquad m\,, n\ge 0\,,\nn\\
{[} P^i_m,P^j_n{]} &=& 0\,,\qquad m\ge0 >n\,,\nn\\
{[} P^i_m,P^j_n{]} &=& f^{ij}{}_k\, (P^k_{m+n} -P^k_m-P^k_n)\,,\qquad
  m\,, n <0\,.\label{PQredef}
\eea
Thus the non-negative modes of $P^i_m$ generate a half Kac-Moody algebra 
that commutes with the negative modes.  The negative modes themselves
generate an algebra that is ``nearly'' another half Kac-Moody algebra.  

   This can be made more precise by expressing the generators $Q^i_m$ of
the algebra obtained in \cite{avabab} in terms of the 
$\hat G\times \hat G$ generators
$J^i_m$ and $\bar J^i_m$ that we introduced in (\ref{KMmodes}) and 
(\ref{KKcom}).  Specifically, the algebra (\ref{avababalg}) found
in \cite{avabab} is obtained if we define
%%%%
\bea
Q^i_m &=& J^i_m\,,\qquad m\ge0\,,\nn\\
Q^i_m &=& -\bar J^i_m + J^i_0 + \bar J^i_0\,,\qquad m\le 0\,.
\eea
%%%%%
In terms of the redefined generators $P^i_m$ of (\ref{PQredef}), we then have
%%%%%
\bea
P^i_m&=& J^i_m\,,\qquad m\ge 0\,,\nn\\
P^i_m&=& \bar J^i_m-\bar J^i_0\,, \qquad m<0\,.
\eea
%%%%%
As well as making manifest that the generators $P^i_m$ for $m\ge0$ commute
with those for $m<0$, it also shows that the algebra (\ref{avababalg}) of
\cite{avabab} is a subalgebra of $\hat G_L\times \hat G_R$.  Furthermore,
it can be seen that if the generators $\bar J_0^i$ were included also,
the algebra would be precisely two commuting copies of a half Kac-Moody
algebra.\footnote{In fact the generators $\bar J^i_0$ must certainly give 
symmetries of the PCM, since they just correspond to the action of the
right-handed factor in the manifest $G_L\times G_R$ Lie algebra symmetry
of the PCM.}

\end{document}